\documentclass[12pt,a4paper]{article}
\usepackage[a4paper,left=1.5cm,right=1.5cm,top=1.5cm,bottom=2.5cm,bindingoffset=5mm]{geometry}
\pdfoutput=1
\usepackage{lmodern}
\usepackage[OT1]{fontenc}
\usepackage{hyperref}
\usepackage{adjustbox}
\hypersetup{
	colorlinks   = true, 
	urlcolor     = blue, 
	linkcolor    = blue, 
	citecolor   = red 
}

\usepackage[format=plain,indention=0pt,font=small,labelfont=bf]{caption}
\usepackage{titlesec}

\titleformat{\section}
{\normalfont\fontsize{12}{15}\bfseries}{\thesection}{1em}{}
\titleformat{\subsection}
{\normalfont\fontsize{12}{10}\bfseries}{\thesubsection}{0.8em}{}

\usepackage{mathtools}
\usepackage{physics}
\usepackage{float}
\usepackage{bbm}
\usepackage{amsfonts, amsmath, amssymb, amsthm,graphicx}
\usepackage{etoolbox}

\numberwithin{figure}{section}
\numberwithin{equation}{section}
\usepackage[skip=1pt  plus1pt, indent=20pt]{parskip}

\usepackage[font=scriptsize]{caption}

\newcommand\ddx[1]{\frac{\partial}{\partial #1}}

\DeclareMathOperator\artanh{artanh}
\DeclareMathOperator\arsinh{arsinh}

\newenvironment{multieq}{\begin{equation} \begin{aligned}}{\end{aligned} \end{equation}}

\newenvironment{multieq*}{\begin{equation*} \begin{aligned}}{\end{aligned} \end{equation*}}

\let\oldabstract\abstract
\let\oldendabstract\endabstract
\makeatletter
\renewenvironment{abstract}
{%
	{\list{}{\addtolength{\leftmargin}{0.3em} 
			\listparindent 1em%
			\itemindent    \listparindent%
			\rightmargin   \leftmargin%
			\parsep        \z@ \@plus\p@}%
		\item\relax}%
	{\endlist}%
	\oldabstract}
{\oldendabstract}
\makeatother

\begin{document}
	\title{\Large Entanglement Entropy, Local IR/UV Connection \\and MPS in Weyl-deformed Geometries}
	\author{Leo Shaposhnik\footnote{ \href{mailto:leoshaposhnik@googlemail.com}{leoshaposhnik@googlemail.com}}}
	\maketitle
	\begin{abstract}\noindent We study the behaviour of entanglement entropy in two-dimensional CFTs under Weyl transformations from the Weyl anomaly. Using the Ryu-Takayanagi-formula, we show that these deformations correspond to local deformations of the IR cutoff of AdS$_3$, which gives an example of the local validity of the IR/UV Connection in AdS/CFT. We then use Matrix Product States to demonstrate the behaviour of entanglement entropy under Weyl transformations in the two-dimensional free fermion CFT mapped to the XY-model with site-dependent coupling. We present numerical results for different deformations of the background and discuss issues arising for the standard DMRG procedure for the resulting inhomogeneous interactions due to a nontrivial background.
	\end{abstract}
	\section{Introduction}
	The holographic principle \cite{Susskind95}, relating the degrees of freedom of D+1-dimensional volumes governed by quantum gravity to the degrees of freedom of the D-dimensional boundary, has lead to the finding of an intimate relationship between quantum information and gravity \cite{Snowmass22}.
	The most precise formulation of this connection is given by the observation of Ryu and Takayanagi \cite{ryu06}, that the entanglement entropy of subsystems of the boundary is related to the area of certain codimension-2 hypersurfaces in the bulk. Studies of these relations in the context of AdS/CFT \cite{Maldacena99} have led to recent progress in the understanding of black holes and unitarity of their evolution \cite{Almheiri19}, which is considered to be a key problem in modern theoretical physics. These studies have also led to a new perspective on the nature of spacetime itself, with the proposal of it being built up from entanglement \cite{VanRaamsdonk10} or at least the possibility of such a mapping between spacetime geometry and quantum information theoretic quantities encoded on the boundary. This approach, dubbed emergent spacetime, is accompanied by studies, that approach to reconstruct the bulk metric on submanifolds \cite{Engelhardt16}, which in turn are believed to be represented by boundary observables. In this article, we want to fill a previously unmentioned gap in the studies of this relation, namely how local changes of the entanglement in the boundary can be mapped to local changes of the radial size of the bulk, in the sense above creating more bulk. To realize this, we study two-dimensional CFTs under Weyl transformations of their background metric 
	\begin{equation}\label{gprime}
		g'_{\mu \nu}(x) = e^{2\sigma(x)} g_{\mu \nu}(x),
	\end{equation}
	to show that these rescaling can be understood as an increase of the radial size of the bulk spacetime.
	This is motivated by the following argument:
	When describing the boundary theory, we regularize it by introducing a cutoff, such as the lattice spacing in a regularization of spacetime by a lattice. In this setting, Weyl transformations act on the proper distance between lattice points as local scale transformations, as noted in \cite{Almheiri19}. As is expected from the IR/UV Connection in AdS/CFT \cite{SusskindWitten98}, this corresponds to a local change of the IR cutoff of the bulk theory, which just is the radial size in our setting.
	As we will verify later, this can be written as
	\begin{equation}\label{drgds}
		\delta \rho(x) = \delta \sigma (x),
	\end{equation}
	where $\delta \rho.\delta \sigma$ are the variation of the IR- and UV cutoff of the bulk and boundary theories.
	In this article, we will demonstrate \eqref{drgds} for the case of AdS$_3$/CFT$_2$.\\
	
	Furthermore, the last three decades have brought another change in the computational methods to study quantum many-body-systems numerically, namely tensor networks (see \cite{TNBook} for a review). Tensor networks are sets of tensors which share indices, that upon contraction of shared indices, can approximate states of a given lattice theory. The individual tensors can be thought of as representing local degrees of freedom in the theory as in the case of Matrix Product States (MPS) or as in the case of MERA, giving a discrete realization of the holographic principle (see \cite{Jahn_2021} for a short review). By reducing the dimensionality of indices shared by tensors, one reduces the dimensionality of subspaces represented by those tensors considered and simplifies the problem of finding e.g. a groundstate of a given system. As we will see, the above considerations provide a simple means to use tensor network methods to problems in quantum field theory in curved spacetime. For constant time slices in AdS$_3$ with a finite but large cutoff, the dual theory will live on the boundary, which is a circle. As we will see below, the behaviour of entanglement entropy under Weyl transformations is universal across CFTs, which allows us to study it with any CFT, not just ones with AdS as the holographic dual. Since the circle is a compact space, the entanglement entropy, being a continuous function of the boundary of the subsystem is bounded. In such situations, Matrix Product States can well represent the CFT-groundstate, provided one employs high enough bond-dimensions\footnote{We thank Michal. P. Heller for pointing this out.}. We will use this situation, to study the above deformations and their effect on the entanglement entropy of the CFT groundstate explicitly by approximating the ground-state of a CFT in different background metrics. This provides a new method for the study of metric perturbations via Weyl transformations, for earlier studies see \cite{tonni18} and references therein.\\
	
	In Section 2, we will derive the behaviour of entanglement entropy under Weyl-rescalings from the Trace anomaly in the boundary theory, a derivation that we could not find in the literature, besides being relatively simple. In Section 3 we will then derive \eqref{drgds} by usage of the Ryu-Takayanagi-formula \cite{ryu06}. This provides an explicit realization of the IR/UV Connection. In Section 4, we present numerical results from DMRG-calculations with MPS, demonstrating the applicability of MPS to investigations of entanglement entropy in CFTs in curved backgrounds and providing a simple tool, with which these questions can be studied with very little computational power.
	\section{Entanglement Perturbations from the Weyl Anomaly}
	In the following section, we derive how the entanglement entropy of the groundstate of a two-dimensional CFT behaves under Weyl rescaling. Our derivation was motivated by a remark in \cite{nishioka19}, which solved the problem for flat spacetime. We give this derivation as an entry point and because we found it to be considerably simpler than earlier derivations, such as in (\cite{tonni18},\cite{Almheiri19}), as well as emphasizing the connection between the quantum mechanical stress tensor and gravity. The four-dimensional case was treated in \cite{solodukhin08} similarly with the inclusion of further complications such as the extrinsic geometry of the entangling surface.
	To begin, we remind the reader of the replica method, used to compute entanglement entropies in quantum field theories. \\
	
	The CFT in question lives on a two-dimensional manifold  $\mathcal{R}$ with Cauchy slices $\Sigma$ and we consider a subsystem $I \subset \Sigma$. $\mathcal{R}$ assumed to be related to the plane $\mathbb{R}^2$ by a conformal transformation, such as the cylinder $\mathbb{R} \times S^1$.
	As was demonstrated in \cite{cardycalabrese04}, the entanglement entropy of $I$ in the CFT-groundstate can be computed by the following formula
	\begin{multieq}\label{edef}
		S_I = -\lim_{n \rightarrow 1}\ddx{n} \Big[\log(Z_n)-n\log(Z)\Big],
	\end{multieq}%
	where $Z_n$ is the partition function of the CFT living on an n-sheeted version of $\mathcal{R}$ that has branch-cuts between the sheets along the subsystem $I$ and $Z$ is the partition function of the CFT.\\
	
	The corresponding path integral can be written as
	\begin{multieq}\label{Zn}
		Z_n = \int [\text{D}\phi] e^{- S_n[\phi]},
	\end{multieq}%
	where $S_n$ represents the action functional on the n-sheeted Riemann surface. The partition function $Z$ can be obtained from $Z_n$ by taking $n=1$. The corresponding action will henceforth be denoted by $S$.\\ 
	
	To learn how the entanglement entropy changes under Weyl rescalings, we compute the first variation of \eqref{edef} under \eqref{gprime}.
	We find
	\begin{multieq}
		\delta S = \lim_{n \rightarrow 1 } &-\ddx{n} \Big[  \frac{1}{Z_n} \int [\text{D}\phi]e^{- S_n[\phi]}
		\ \delta S_{n}[\phi] - \frac{n}{Z}\int [\text{D}\phi] e^{- S[\phi]} \delta S[\phi] \Big].
	\end{multieq}%
	Using the definition of the stress-energy tensor
	\begin{equation}\label{emtensor}
		T^{\mu\nu}(x) = \frac{4\pi}{\sqrt{g}}\frac{\delta S}{\delta g_{\mu\nu}}(x),
	\end{equation}
	together with the infinitesimal Weyl rescaling $\delta g_{\mu\nu}(x) = 2 \delta \sigma(x) g_{\mu\nu}(x)$, we obtain
	\begin{multieq}
		\delta S_I &= \lim_{n \rightarrow 1 }\ddx{n} \Big[  \frac{1}{Z_n} \int [\text{D}\phi] e^{- S_n[\phi]}
		(\int_{\mathcal{R}_n}\text{d}^2x \frac{\sqrt{g}}{2\pi}T_{\mu}^{\mu}\delta \sigma(x)) - \frac{n}{Z}\int [\text{D}\phi] \int_{\mathcal{R}}\text{d}^2x \frac{\sqrt{g}}{2\pi}T_{\mu}^{\mu}\delta \sigma(x) \Big]\\
		&=\lim_{n \rightarrow 1 }\ddx{n} \Big[   (\int_{\mathcal{R}_n}\text{d}^2x \frac{\sqrt{g}}{2\pi}\langle T_{\mu}^{\mu}\rangle_{\mathcal{R}_n} \delta \sigma(x)) - n (\int_{\mathcal{R}}\text{d}^2x \frac{\sqrt{g}}{2\pi}\langle T_{\mu}^{\mu}\rangle_{\mathcal{R}}\delta \sigma(x)) \Big],
	\end{multieq}%
	where $\mathcal{R}_n$ denotes the n-sheeted Riemann surface.
	Now we follow the review \cite{nishioka19} and assume that the Weyl anomaly persists on the n-sheeted surface. In particular, this means
	\begin{multieq}
		\langle T \rangle_{\mathcal{R}_n}(x) &= -\frac{c}{12} R(x) \biggr \rvert_{\mathcal{R}_n},
	\end{multieq}%
	where $R(x)$ denotes the Ricci scalar. This assumption is natural because the trace is a scalar function and since the two surfaces are connected by a holomorphic transformation, the two expectation values are expected to be the same.
	It was shown in \cite{fursaev95}, that the curvature on the n-sheeted surface is given by
	\begin{multieq}\label{Rrn}
		\sqrt{g}R(x)|_{\mathcal{R}_n} = \sqrt{g}R|_{\mathcal{R}}+4 \pi (1-n) \delta_{\Sigma} + \mathcal{O}((1-n)^2),
	\end{multieq}%
	where $\delta_{\Sigma}$ is a delta function for the entangling-surface
	\begin{multieq}
		\delta_{\Sigma} = \delta(x-u) + \delta(x-v),
	\end{multieq}%
	if the subsystem lives on the interval is $I = [u,v]$.	Combining everything and neglecting terms of higher order in $(1-n)$, we obtain
	\begin{multieq}\label{dsI}
		\delta S_I &=-\frac{c}{24\pi}\lim_{n \rightarrow 1 } \ddx{n} \Big[   (\int_{\mathcal{R}_n}\text{d}^2x \sqrt{g}( R|_{\mathcal{R}}+
		4 \pi (1-n) \delta_{\Sigma})\delta \sigma(x)) - n (\int_{\mathcal{R}}\text{d}^2x \sqrt{g}R|_{\mathcal{R}}\delta \sigma(x)) \Big]\\
		&=\frac{c}{6}(\delta \sigma(u)+\delta \sigma(v)) -\frac{c}{24\pi} \ddx{n} \delta \tilde{Z}_n ,
	\end{multieq}%
	where we defined
	\begin{multieq}
		\delta \tilde{Z}_n := \int_{\mathcal{R}_n}\text{d}^2x \sqrt{g} R|_{\mathcal{R}}\delta \sigma(x) - n \int_{\mathcal{R}}\text{d}^2x \sqrt{g}R|_{\mathcal{R}}\delta \sigma(x).
	\end{multieq}%
	This contribution vanishes because the two integrals differ at most by the endpoints of the interval, which form a set of measure zero.
	Thus we find the following result for perturbations of the entanglement entropy under Weyl transformations
	\begin{multieq}\label{dS}
		\delta S_I =\frac{c}{6}(\delta \sigma(u)+\delta \sigma(v)).
	\end{multieq}%
	The full entanglement entropy in a Weyl-rescaled geometry  $g' = e^{2\sigma}g$ is then found by integrating \eqref{dS}
	\begin{multieq}\label{Sfull}
		S_{g'} = S_{g} + \frac{c}{6}(\sigma(u)+\sigma(v)),
	\end{multieq}%
	where $S_{g}$ is the entanglement entropy in the background $g$. Later on, we will consider a flat background $g$. Since the above derivation only employed the Weyl anomaly being the same in all CFTs, it holds in all two-dimensional CFTs.
	\section{Local IR/UV Connection in AdS$_3$/CFT$_2$}
	We now come to the derivation of \eqref{Sfull} from the geometry dual to the CFT-groundstate, given by empty AdS$_3$. Our derivation is inspired by the original paper of Ryu and Takayanagi \cite{ryu06}, that introduces their conjecture relating the entanglement entropy $S_I$ of the system $I$ of the CFT to the area of a minimal surface $\gamma$, which connects the boundary points of $I$ and passes through the bulk. Schematically this is given by
	\begin{multieq}\label{RTC}
		S_I = \min_{\gamma:\ \partial \gamma = \partial I} \frac{\mathcal{A}(\gamma)}{4G},
	\end{multieq}%
	where $\mathcal{A}(\gamma)$ denotes the area of the surface $\gamma$ and we take the minimum over all surfaces homologous to I. $G$ is Newtons constant of the bulk-spacetime. We will only consider stationary systems, i.e. time-independent metrics, where \eqref{RTC} suffices.\\
	
	In \cite{ryu06}, the authors consider a constant time-slice of AdS$_3$ in global coordinates, described by the metric
	\begin{multieq}\label{ads}
		ds^2= R^2(-\cosh \rho^2 \text{d}t^2 + \text{d}\rho^2 + \sinh^2 \rho \text{d}\theta^2),
	\end{multieq}%
	where the coordinates $\rho, \theta$ take the values
	\begin{multieq}
		0 < \rho \leq \rho_0,\ -\infty < t < \infty,\ 0 \leq \theta  < 2 \pi.
	\end{multieq}%
	Here an IR cutoff $\rho_0$ was introduced.
	To introduce a curved boundary, we perform the following modification, as suggested by the argument following \eqref{drgds}:\\
	
	Instead of considering a fixed cutoff $\rho_0$, we introduce an angle-dependent cutoff $\rho_0(\theta)$. 
	For the sake of transparency, we will describe the computation in some detail.\\
	
	As stated by \eqref{RTC}, to find the entanglement entropy of the subsystem, we need to find the minimal surface connecting the boundary points of the subsystem $I$. We set $I$ to be given by the circle element with boundary points $(\rho_f,\theta_f),(\rho_i,\theta_i)$, where $\rho_{f,i} = \rho(\theta_{f,i})$.
	As described in \cite{louko00}, the minimal surface connecting the boundary points is given by the curves
	\begin{multieq}\label{geodesic}
		\tanh \rho(\theta) \cos (\theta - \theta_B) = \cos(\alpha).
	\end{multieq}
	Here $\theta_B, \alpha$ are integration constants needed to satisfy the boundary conditions
	\begin{multieq}\label{constraints}
		\tanh \rho_i \cos (\theta_i - \theta_B) &= \cos(\alpha),\\
		\tanh \rho_f \cos (\theta_f - \theta_B) &= \cos(\alpha).
	\end{multieq}%
	In the regime of interest $\rho_{f,i} \gg 1$, the hyperbolic tangens becomes 1 and the constant $\theta_B$ reduces to 
	\begin{multieq}\label{thetab}
		\theta_B = \frac{\theta_f + \theta_i}{2}.	
	\end{multieq}%
	Combining \eqref{ads} and \eqref{geodesic} the length $A$ of such a geodesic is then given by
	\begin{multieq}\label{length}
		&A(\theta_f,\theta_i) = R \int_{\theta_i}^{\theta_f}\text{d}\theta \sqrt{\frac{d \rho}{d \theta} + \sinh^2 \rho}
		=  R [\artanh(\frac{\tan \theta_f-\theta_B}{\tan \alpha})-\artanh(\frac{\tan \theta_i-\theta_B}{\tan \alpha})].
	\end{multieq}%
	Using \eqref{constraints} one can derive the identity
	\begin{multieq}\label{Id1}
		\frac{\tan (\theta-\theta_B)}{\tan \alpha} &= \frac{\sin (\theta- \theta_B)\sinh(\rho)}{\sqrt{1+ \sinh^2(\rho)\sin^2(\theta-\theta_B)}}.
	\end{multieq}%
	Combining \eqref{length},\eqref{Id1},\eqref{constraints} and
	defining the function $F(\theta) = \sin(\theta - \theta_B)\sinh(\rho(\theta))$,  we obtain
	\begin{equation}
		A(\theta_f,\theta_i) = \frac{R}{2}[\arsinh(F(\theta_f))-\arsinh(-F(\theta_f))-\arsinh(F(\theta_i))+\arsinh(-F(\theta_i))].
	\end{equation}
	With the identity $\arsinh(-x) = -\arsinh(x)$ we find\texttt{}
	\begin{multieq}\label{EEcft}
		A(\theta_f,\theta_i) = R [\arsinh(F(\theta_f))-\arsinh(F(\theta_i))].
	\end{multieq}%
	In the regime $\rho_0 \gg 1$, we have
	\begin{equation}\label{asymptotic}
		F(\theta_{f,i}) \sim \pm e^{\rho_0 +\delta \rho_{i,f}} \sin(\frac{\theta_f-\theta_i}{2}),
	\end{equation} 
	where the $+$($-$) sign holds for $f$($i$). Combining \eqref{asymptotic} with \eqref{EEcft} and using
	\begin{equation}
		\lim_{x\rightarrow \infty }\frac{\arsinh(x)}{\ln(2x)}=1,
	\end{equation} 
we find
	\begin{equation}
		A(\theta_f,\theta_i) \rightarrow R \ln(\sin^2(\frac{\theta_f-\theta_i}{2})e^{2\rho_0})+ R ( \delta \rho(\theta_f) + \delta \rho(\theta_i)).
	\end{equation}
	Using the Ryu-Takayanagi-formula \eqref{RTC}, we obtain the central charge 
	\begin{equation}
		c = \frac{3R}{2 G_n^{(3)}},
	\end{equation}
	which matches the derivation from \cite{ryu06} for a flat boundary.\\
	
	A comparison with \eqref{Sfull} indeed gives the relation \eqref{drgds}, after we identify the flat-spacetime result
	\begin{equation}
		A_{flat}(\theta_f,\theta_i) = R\ln(\sin^2(\frac{\theta_f-\theta_i}{2})e^{2\rho_0}).
\end{equation}
	We see that the Weyl-rescaled boundary metric can be described by an angle-dependent IR cutoff in the bulk theory.	Note that this is an identification of dimensionless parameters, since in our choice of coordinates \eqref{ads}, $\rho$ is dimensionless.	
	
	\section{Entanglement Perturbations from MPS in curved Spacetime}
	We will now present, how to treat the above considerations on the CFT side numerically by using Matrix Product States (MPS). As we noted at the end of section 2 the result holds for the groundstate of any CFT, not just ones with a holographic dual. We construct a spin chain for free, two-dimensional fermions in a Weyl-deformed geometry, by using the Jordan-Wigner transformation \cite{JordanWigner28} on the staggered fermion description given in \cite{lewis20} and compute the entanglement entropy with standard tensor network techniques. This spin chain turns out to be the XY-model with site-dependent coupling, which was also obtained by different methods, as a discrete model for free fermions in curved spacetime, see (\cite{yang20}, \cite{tonni18}) for earlier studies.\\
	
	In \cite{lewis20}, a discretization was found for the two-dimensional free fermion CFT with central charge $c=1$, given by the action
	\begin{multieq}\label{lagrangian}
		S &= -\frac{1}{2} \int \text{d}^2 \mathcal{L} \sqrt{-g}(\bar{\psi}\tilde{\gamma}^{\mu}D_{\mu}\psi-D_{\mu}\bar{\psi}\tilde{\gamma}^{\mu}\psi),
	\end{multieq}%
	where $\psi$ is a two-dimensional spinor and $D_{\mu}$ the covariant Spinor derivative. The lattice model on $N$ points is derived by replacing the field $\psi$ by staggered fermions \cite{Susskind77} $\phi_n, n \in [1,2,\hdots,N]$, that satisfy canonical anticommutation relations $ \{\phi_n,\phi^{\dagger}_m\} = \delta_{nm}$ and the following Hamiltonian is found
	\begin{equation}
		H_{\text{lattice}} = \sum_{n=1}^N-i\frac{\Omega(x_n)}{d}
		\big[
		\phi^{\dagger}_n\phi_{n+1}-\phi^{\dagger}_{n+1}\phi_{n}
		\big].
	\end{equation}
We choose periodic boundary conditions $\phi_{N+1} = \phi_{1}$, to approximate the CFT on a cylinder, which is the situation on the boundary of AdS$_3$.
	We have chosen a constant proper distance $d > 0$, such that the Weyl factor $\Omega(x_n)$ appears in the Hamiltonian. If one instead chooses a constant coordinate distance $a$, as was done in \cite{lewis20}, one obtains the same Hamiltonian with the substitution $\frac{\Omega(x_n)}{d} \rightarrow \frac{1}{a}$.\\
	
	Now we follow \cite{Susskind77} and perform a Jordan-Wigner transformation\cite{JordanWigner28}
	\begin{multieq}
		\phi_n^{\dagger} &= \prod_{m < n} i \sigma_m^z \sigma_n^+,\\
		\phi_n &= \prod_{m < n} (-i \sigma_m^z) \sigma_n^-,
	\end{multieq}%
	where $\sigma^{x,y,z}$ are the usual Pauli-matrices and $\sigma^{\pm} = \frac{1}{2}(\sigma^x \pm  i \sigma^y)$. This leads us to the following Hamiltonian 
	\begin{equation}\label{hxx}
		H_{xy} =  \sum_n\frac{\Omega(x_n)}{2d}
		\big[\sigma_n^x \sigma_n^x+
		\sigma_n^y\sigma_{n+1}^y
		\big],
	\end{equation}
	which is the Hamiltonian of the XY-model with site-dependent coupling. This was previously derived by \cite{yang20} and has earlier been noted by \cite{tonni18} and references therein, the derivation in terms of staggered fermions appears to be the most straightforward method to come to this discretization.\\
	
	Here the interpretation from the previous section of the Weyl factor arising from a position-dependent IR cutoff becomes visible by the now spatially dependent coordinate UV cutoff $\frac{\Omega(x)}{d}$.
	Since \eqref{hxx} is a spin chain representing a 2d-CFT on a compact manifold, which leads to bounded entanglement, entropy, we have reason to believe, that we can approximate its groundstate by a Matrix Product state \cite{Schuch08}.
	To demonstrate this, we take chains of sizes $N \in \{2,4,6,8,10\}$ and compute its groundstate via DMRG (\cite{White92},\cite{White93},\cite{RommerOstlund97}). For uneven $N$, we do not expect the system to behave as the fermion theory because the staggered fermions used in the derivation of the model, are based on a doubling of the lattice sites, to accompany the two fermionic degrees of freedom in two dimensions. This mapping is not consistent for an uneven number of lattice sites.\\

	\subsection{A first Weyl Factor}
	To get started, we choose Weyl factors of the form
	\begin{equation}
		\Omega(x) = e^{\kappa \sin(pi x)},
	\end{equation} 
with $\kappa \in\{-1,0,1\}$ to have both negative and positive curvature and respect the periodicity and $x \in [0,1)$ is the coordinate used to parametrize the circle on which the CFT lives.
	We stopped the DMRG calculation when the variance of the energy satisfied 
	\begin{equation}
		 \langle H^2\rangle-\langle H \rangle ^2 \leq 10^{-12},
	\end{equation} 
which takes a few minutes on a normal laptop for the system sizes we consider.\\

 After obtaining a MPS, approximating the groundstate of the spin-chain, we computed its entanglement entropy $S(x)$ and subtracted a normalizing constant $C$, such that the value at the point $\bar{x}$ closest to the centre of the circle coincided with the theoretical prediction. This was chosen because this point represents the largest subsystem possible and the CFT should be approximated better with increasing systemsize. Here, $x \in [0,1)$ labels the coordinate of the cylinder. We then computed the entanglement perturbations given in \eqref{s_perturbations}. We show the results in Fig. \ref{s_plot}.
	\begin{multieq}\label{s_perturbations}
	\bar{S}(x) &= S(x)-C,\ C = S(\bar{x})-S_{flat}( \bar{x} ),\\
	\delta \bar{S}(x) &= S(x)-C-S_{flat}(x).\\
	\end{multieq}%
We see that already for small systems of size $N=10$ we have very good agreement between numerical results and theory. Note that for positive $k$, positive curvature, the agreement is worse than for negative $k$.
To investigate the convergence to the continuum limit, we computed the variance $\sigma(\delta^2\bar{S})$ for increasing systemsizes, where
\begin{multieq}
	\delta^2\bar{S}(x) &= S(x)-C-S_{flat}(x)-\frac{1}{6}(\delta \sigma(0)+\delta \sigma(x)),\\
	\sigma(\delta^2\bar{S}) &= (\frac{1}{N-1}\sum_{i=1}^N(\delta^2\bar{S} - \langle \delta^2 \bar{S}\rangle)^2)^{1/2},
\end{multieq}%
are the difference between the numerical result to the theoretical prediction and its empirical variance. We have set $c=1$ to match the free fermion CFT. Here $\langle \delta^2 \bar{S}\rangle$ denotes the mean over the chain. We computed the groundstate for systemsizes $ N \in \{4,\hdots,20\}$ and show the results in Fig. \ref{fig_conv}. We see that for $\kappa = 1$ the entanglement entropy of the resulting MPS converges considerably slower than for $\kappa = 0,-1$.
 We suspect that this originates in the nonperiodic nature of the MPS tensor networks we chose. The two hamiltonians \eqref{hxx} with  $\kappa-1, \kappa=1$ are identified by a shift $x \rightarrow x+1$, thus we would expect the same numerical stability. In the approximation via MPS, the largest conformal factor is present in the centre of the chain for $\kappa=1$ and for $\kappa-1$ the smallest conformal factor. Thus for $\kappa=1$ we have increasing interaction between spins going towards the centre, needing increasingly large bond dimensions to capture the correlations. We leave it for future studies to investigate this issue in detail.
 \begin{figure}[H]\label{s_plot}
 	\hspace{2,7cm}
 	\includegraphics[scale=0.4]{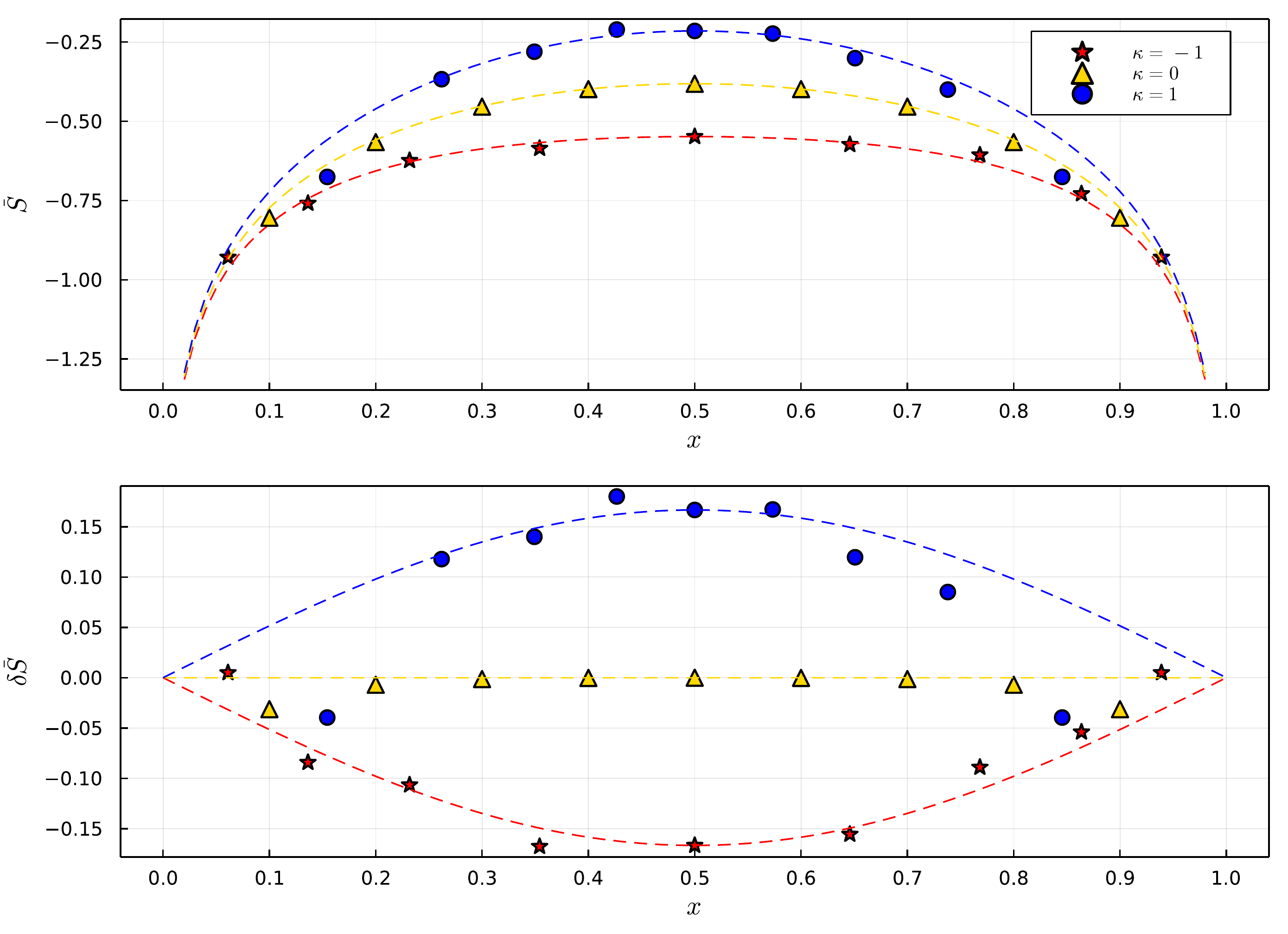}
 	\captionsetup{width=.8\linewidth}
 	\caption{Entanglement entropy and its perturbations for the groundstate of the XY-model with Weyl factors $\Omega(x) = e^{\kappa \sin(\pi x)}$ for $N=10$ points. Dashed lines describe the theoretical prediction.}
 	\label{s_plot}
 \end{figure}
 \begin{figure}[H]\label{fig_conv}
 	\hspace{4cm}
 	\includegraphics[scale=0.4]{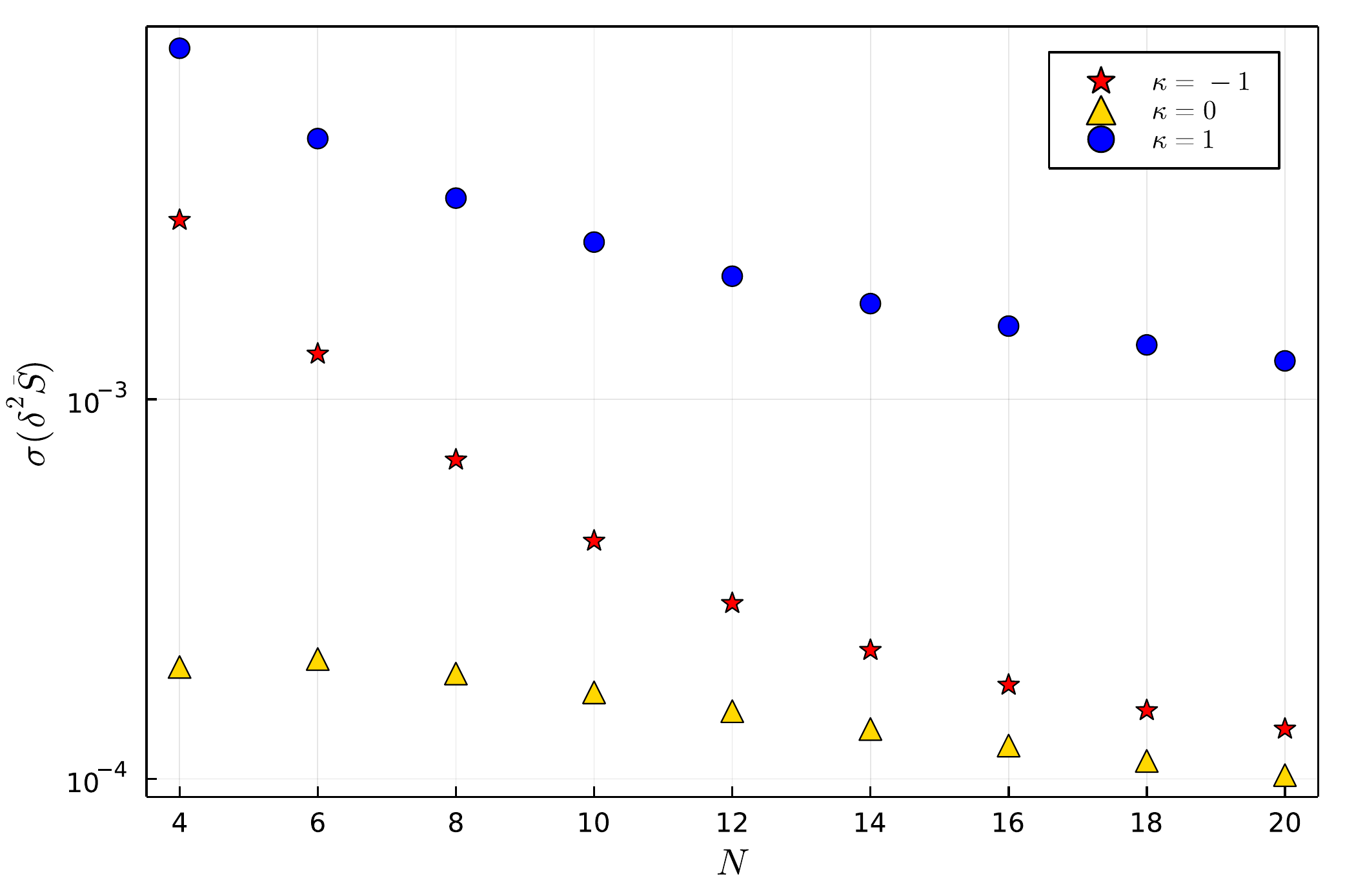}
 	\captionsetup{width=.8\linewidth}
 	\caption{ Variance of $\delta^2\bar{S}$ for increasing systemsizes for Weyl factors $\Omega(x) = e^{\kappa \sin(\pi x)}$.}
 	\label{fig_conv}
 \end{figure}
\subsection{A second Weyl Factor}
The previous discussion was based on Weyl factors of the form $\Omega(x) = e^{\kappa \sin(\pi x)}$, which give a very simple discretization. As is visible from Fig. \ref{s_plot} above, the lattice is spread evenly across the circle and thus different regions of the circle are probed with a similar number of points. This changes if one uses Weyl factors that have an increased variation, where due to our construction using a lattice of constant proper distance between neighbouring sites, the concentration of lattice points can differ drastically if the background in different regions has for example opposite curvature. To test this, we now consider backgrounds with
\begin{equation}
	\sigma(x) = \kappa \sin(2\pi x),\ \kappa \in\{-1,1\}.
\end{equation}
We show the results in Figs. \ref{s_plot_2}, \ref{fig_conv_2}. In Fig. \ref{s_plot_2} we show a chain of size $N=20$, since we need to consider chains with more points because as is visible from Fig. \ref{s_plot_2}, the region where $\sigma$ becomes negative is covered by only a small number of points. We still see that for relatively small systems of size $N=20$ we obtain good agreement between the theoretical prediction and numerical results.\\

A surprising point is that in principle, both prefactors with $\kappa = 1$ and $\kappa =-1$ are physically the same. but we see in Fig. \ref{fig_conv_2}, that their convergence differs and also in Fig. \ref{s_plot_2} it is visible, that the MPS approximates the entanglement structure better for $\kappa=1$.\\

 We suspect that this comes from the way DMRG is performed, by solving a local minimization problem starting from the left (the sites at $x=0$) and going to the right and then back and forth. By starting on the left for $\kappa=1$ the first approximations of the groundstate are done in the region in which most of the points reside and interact strongly, which influences the rest of the approximation. In this case, since the part of the chain that interacts the most is approximated first, we expect better results than for an approximation that follows the opposite direction. This is exactly the case for $\kappa =-1$, where the beginning of the chain interacts weakly in opposite to the end. This hints towards the need for a different updating procedure for inhomogeneous interactions in general, where one would update the sites interacting strongly first, and then move towards the sites with weaker interactions or the use of manifestly periodic Matrix Product States.

 \begin{figure}[H]\label{s_plot_2}
	\hspace{2.7cm}
	\includegraphics[scale=0.4]{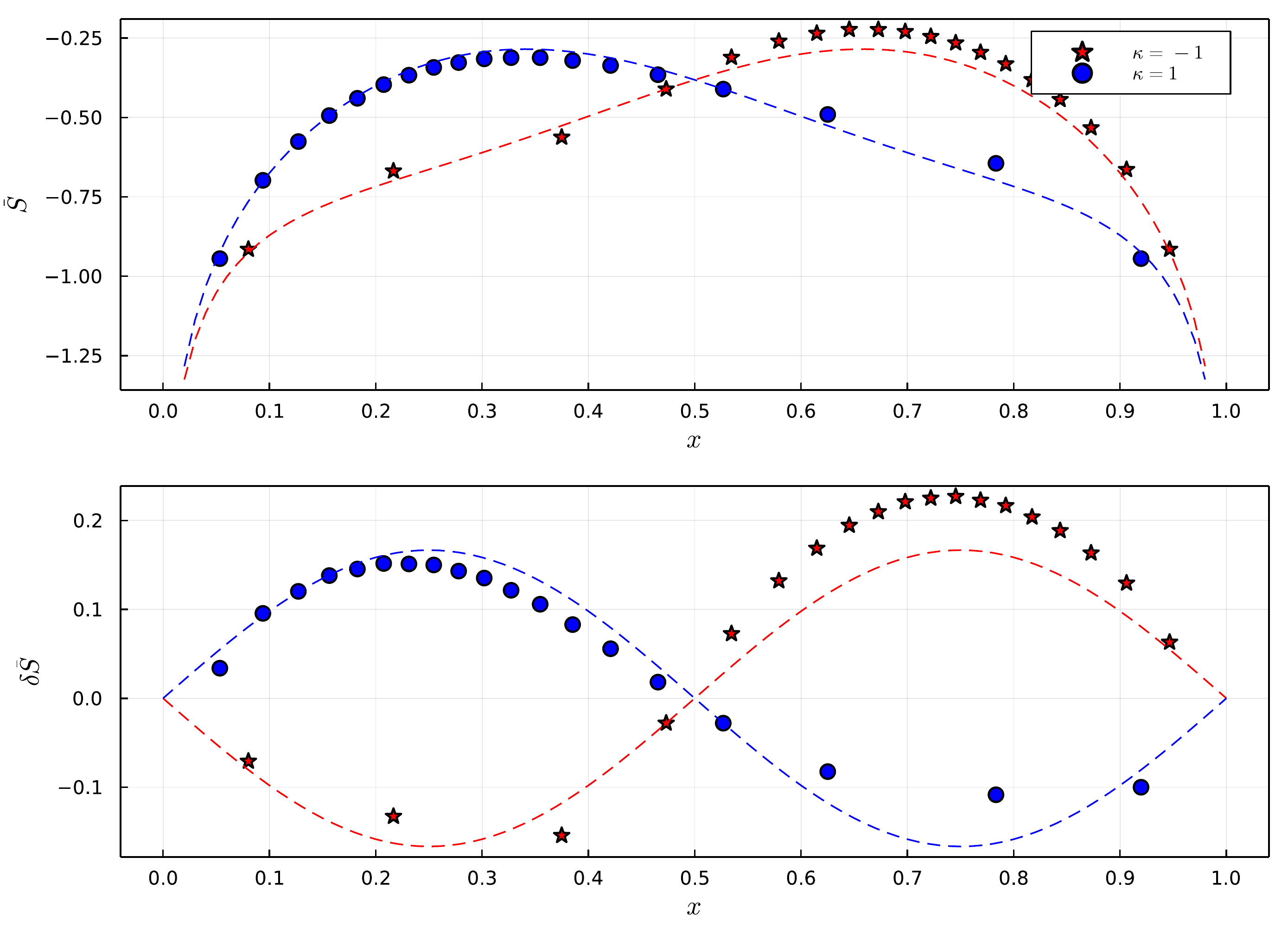}
	\captionsetup{width=.8\linewidth}
	\caption{Entanglement entropy and its perturbations for the groundstate of the XY-model with Weyl factors $\Omega(x) = e^{\kappa \sin( 2 \pi x)}$ for $N=20$ points. Dashed lines describe the theoretical prediction.}
	\label{s_plot_2}
\end{figure}
\begin{figure}[H]\label{fig_conv_2}
	\hspace{4cm}
	\includegraphics[scale=0.4]{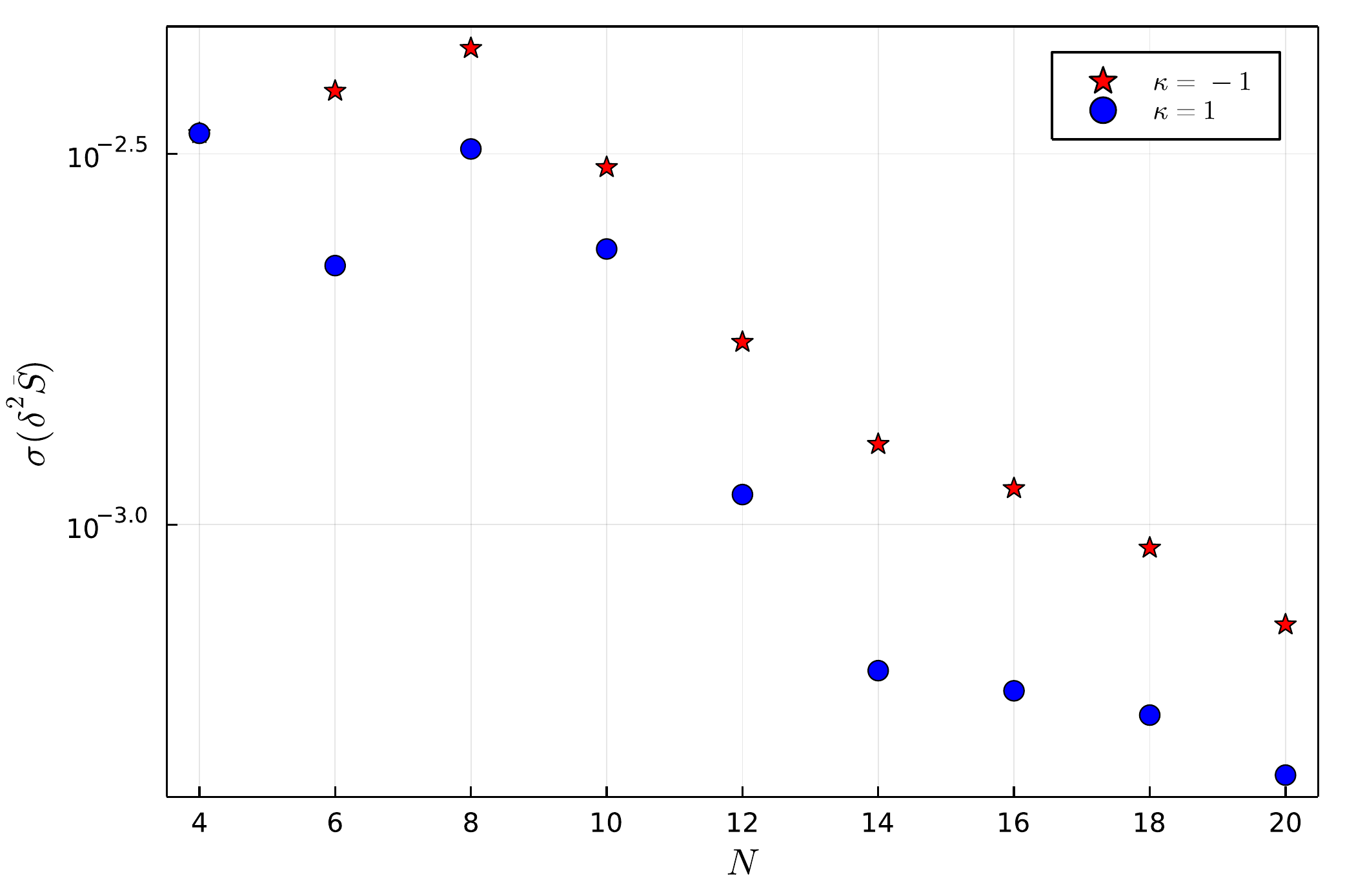}
	\captionsetup{width=.8\linewidth}
	\caption{ Variance of $\delta^2\bar{S}$ for increasing systemsizes for Weyl factors $\Omega(x) = e^{\kappa \sin( 2 \pi x)}$.}
	\label{fig_conv_2}
\end{figure}
  \section{Discussion}
  	The study of quantum information theoretic quantities in the context of the holographic duality has been a fruitful path to understand different aspects of quantum gravity. Here, we have given a local instantiation of the IR/UV Connection which was previously undescribed, relating local deformations of the boundary UV cutoff to local deformations of the bulk IR cutoff. Previous discussions of the IR/UV Connection focused on global deformation of the cutoff and in that regard, we could fill a gap. Although the result for the entanglement deformations \eqref{dS} is well known in the context of two-dimensional conformal field theory and was described in the literature \cite{tonni18} \cite{Almheiri19}, our derivation can be extended straightforward to higher dimensions as was done in \cite{solodukhin08} for the four-dimensional case, which is different from previous derivations in two dimensions, which either relied on the fact that the entanglement entropy of an interval in 2d-CFTs can be written as the two-point function of twist-fields \cite{Almheiri19} or they employed the Liouville action \cite{tonni18} which results from an integration of the Weyl anomaly in two dimensions. Furthermore, our calculation on the AdS$_3$-side is not present in the current literature.\\
  	
  	We further demonstrated the possibility to simulate QFTs in curved spacetimes with Matrix Product States, a demonstration that is to our knowledge the first of its kind. The surprising point is, that to perform these calculations, no large-scale computing facility is needed. The computation of a chain of $N=10$ sites can be done with very little code\footnote{We used the Julia-version of the iTensor library \cite{itensor} to perform our computations, with a program containing about 60 lines.} and a standard office laptop and still reproduces the continuum result reasonably well. We further saw, that the concrete deformation of the background metric strongly influences the convergence. We argued that this is due to the MPS approximation we employed but this point is not clarified completely and is left open for future investigation.\\
  	
  	We then considered more complicated Weyl factors giving both negative and positive deformations of the entanglement entropy and saw that in these systems, we still can reproduce the continuum result well for small systems of $N=20$ sites. We further noted that although the two Weyl factors we considered are physically equivalent, the computations do not give equally good results. We proposed an explanation for this difference, originating in the way the approximation is obtained via DMRG.\\
  	
  	So far we saw no principle obstruction to general Weyl factors as long as the number of sites is large enough to probe all regions of space, which becomes a nontrivial problem for strongly varying Weyl factors. A future perspective would be to look at different ways of discretization, that have varying proper distances but not constant coordinate distance, since as we saw in the discussion below \eqref{hxx}, in this approximation the lattice Hamiltonian becomes the one of the homogeneous XY-model.\\
  	
   \noindent	
   
   \textbf{Acknowledgements:} I thank Michal P. Heller for his supervision of my master's thesis which formed the basis of Sections 2 and 4, reading of the manuscript and helpful comments. I also thank Johannes Knaute for helpful remarks.
  	\newpage
  	
	\bibliography{references}
	\bibliographystyle{h-physrev}
\end{document}